\documentclass[11pt]{article}

\usepackage{asp2006}
\usepackage{graphicx}

\begin{document}
\title{The Cluster-Merger Shock in 1E 0657-56: Faster than a Speeding
  Bullet?}  
\author{Jun Koda\altaffilmark{1}, Milo\v s
  Milosavljevi\'c\altaffilmark{2}, Paul R. Shapiro\altaffilmark{2},
  Daisuke Nagai\altaffilmark{3} and Ehud Nakar\altaffilmark{3} }
\altaffiltext{1}{Department of Physics, University of Texas, 1
  University Station C1600, Austin, TX 78712.}
\altaffiltext{2}{Department of Astronomy, University of Texas, 1
  University Station C1400, Austin, TX 78712.}
\altaffiltext{3}{Theoretical Astrophysics, California Institute of
  Technology, Mail Code 130-33, 1200 East California Boulevard,
  Pasadena, CA 91125.}

\begin{abstract} 
The merging galaxy cluster 1E 0657-56, known as the ``bullet
cluster,'' is one of the hottest clusters known. The X-ray emitting
plasma exhibits bow-shock-like temperature and density jumps. The
segregation of this plasma from the peaks of the mass distribution
determined by gravitational lensing has been interpreted as a direct
proof of collisionless dark matter. If the high shock speed inferred
from the shock jump conditions equals the relative speed of the
merging CDM halos, however, this merger is predicted to be such a rare
event in a $\Lambda$CDM universe that observing it presents a possible
conflict with the $\Lambda$CDM model.

We examined this question using high resolution, 2D simulations of
gas dynamics in cluster collisions to analyze the relative motion of
the clusters, the bow shock, and the contact discontinuity, and relate
these to the X-ray data for the bullet cluster. We find that the
velocity of the fluid shock need not equal the relative velocity of
the CDM components. An illustrative simulation finds that the present
relative velocity of the CDM halos is $16\%$ lower than that of the
shock. While this conclusion is sensitive to the detailed initial mass
and gas density profiles of the colliding clusters, such a decrease of
the inferred halo relative velocity would significantly increase the
likelihood of finding 1E 0657-56 in a $\Lambda$CDM universe.
\end{abstract}



\section{Introduction}
We use gas dynamical simulations of the nearly head-on collision of
two unequal-mass clusters to model the ``bullet cluster'' 1E 0657-56,
to show that the high shock velocity of $4700 \textrm{ km s}^{-1}$
inferred from X-ray observations \citep{Markevitch:06, Markevitch:07}
may exceed the relative velocity of the merging halos. Previous
estimates of the extremely small probability of finding such a
high-velocity cluster-cluster collision in a $\Lambda$CDM universe
assumed that the shock and the halo-collision velocities were the same
\citep{Hayashi:06, Farrar:07}. Our results, first described in
\citet{Milosavljevic:07}, significantly increase the likelihood of
observing such a merger event in $\Lambda$CDM.

\section{Simulation}
We used the adaptive mesh-refinement (AMR) ASC FLASH code
\citep{Fryxell:00} in 2D cylindrical coordinates to simulate a region
$20$ Mpc across, with $1$ kpc spatial resolution.  Gravity is
contributed by two rigid, spherical DM halo profiles, whose centers
free-fall toward each other from rest at separation $4.6$ Mpc. The density
associated with the gravitational force is an NFW
profile $\rho = \rho_0 (r/r_s)^{-1} (1+r/r_s)^{-2}$ for the
subcluster, and a cored profile $\rho = \rho_0 (1+r/r_s)^{-3}$ for the
main cluster, to imitate a non-head-on merger. The masses
($M_{500}/10^{15} M_\odot$) and radii $r_{500}$ and $r_s$ (Mpc) are
$(1.25, 1.5, 0.5)$ and $(0.25, 0.5, 0.16)$ for the main cluster and
subcluster, respectively.

\newlength{\minitwocolumn}
\setlength{\minitwocolumn}{0.25\textwidth}

\begin{figure}[h]
  \begin{tabular}{ccc}
    \begin{minipage}{\minitwocolumn}
      \begin{center}
	\includegraphics[height=6.2cm,clip]{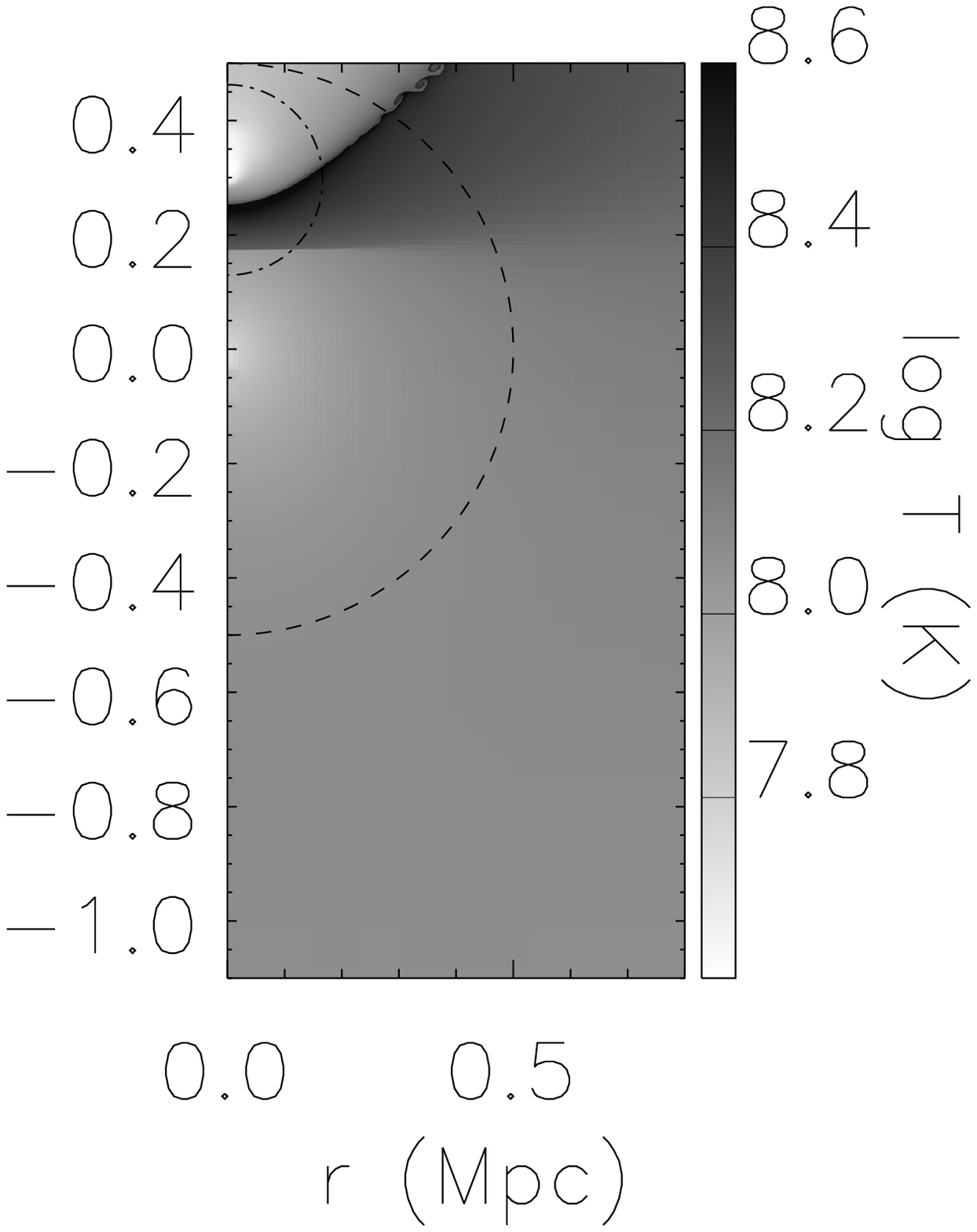}
      \end{center}
    \end{minipage}
    &
    \begin{minipage}{\minitwocolumn}
      \begin{center}
	\includegraphics[height=6.2cm,clip]{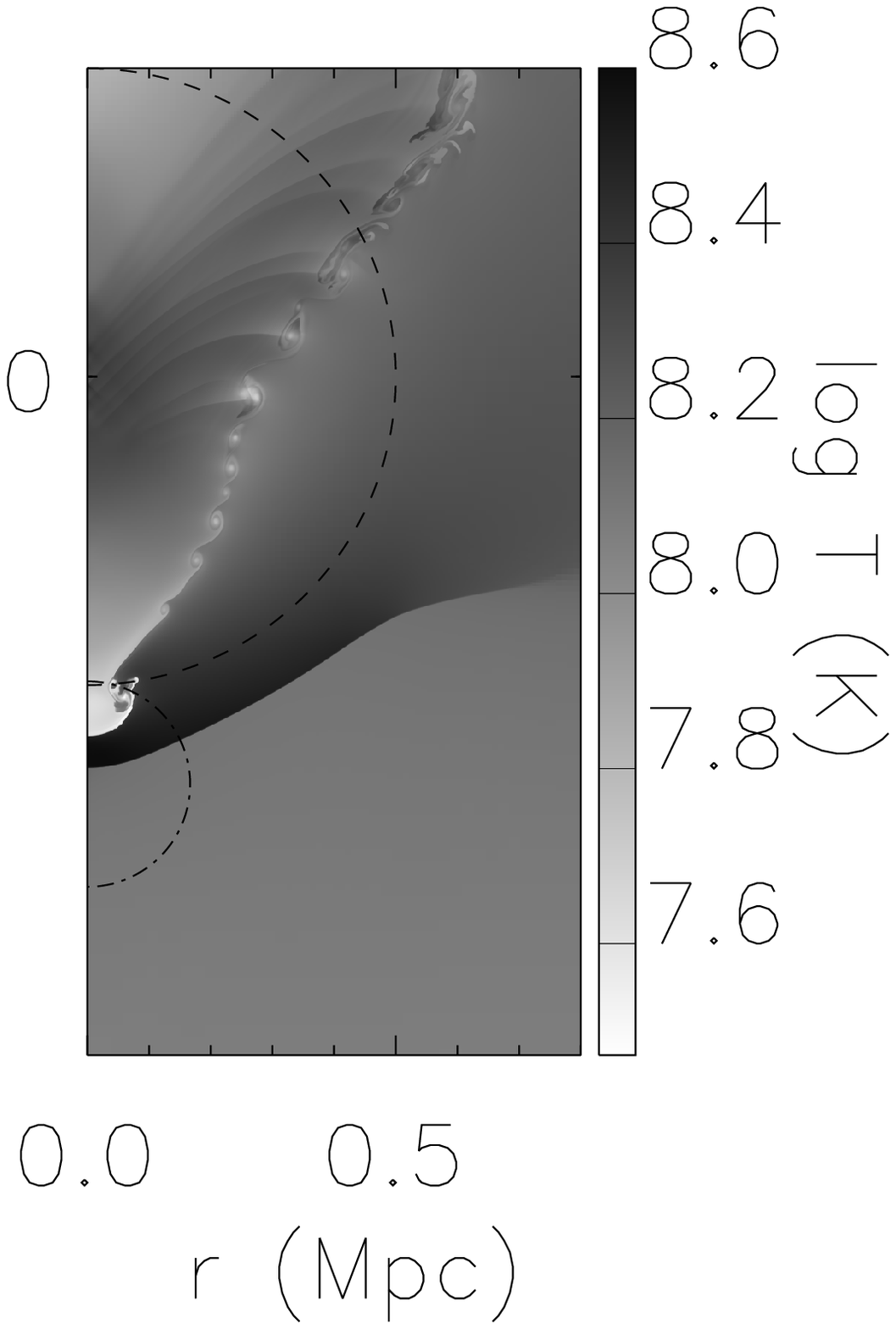}
      \end{center}
    \end{minipage}
    &
    \begin{minipage}{\minitwocolumn}
      \begin{center}
	\includegraphics[height=6.2cm,clip]{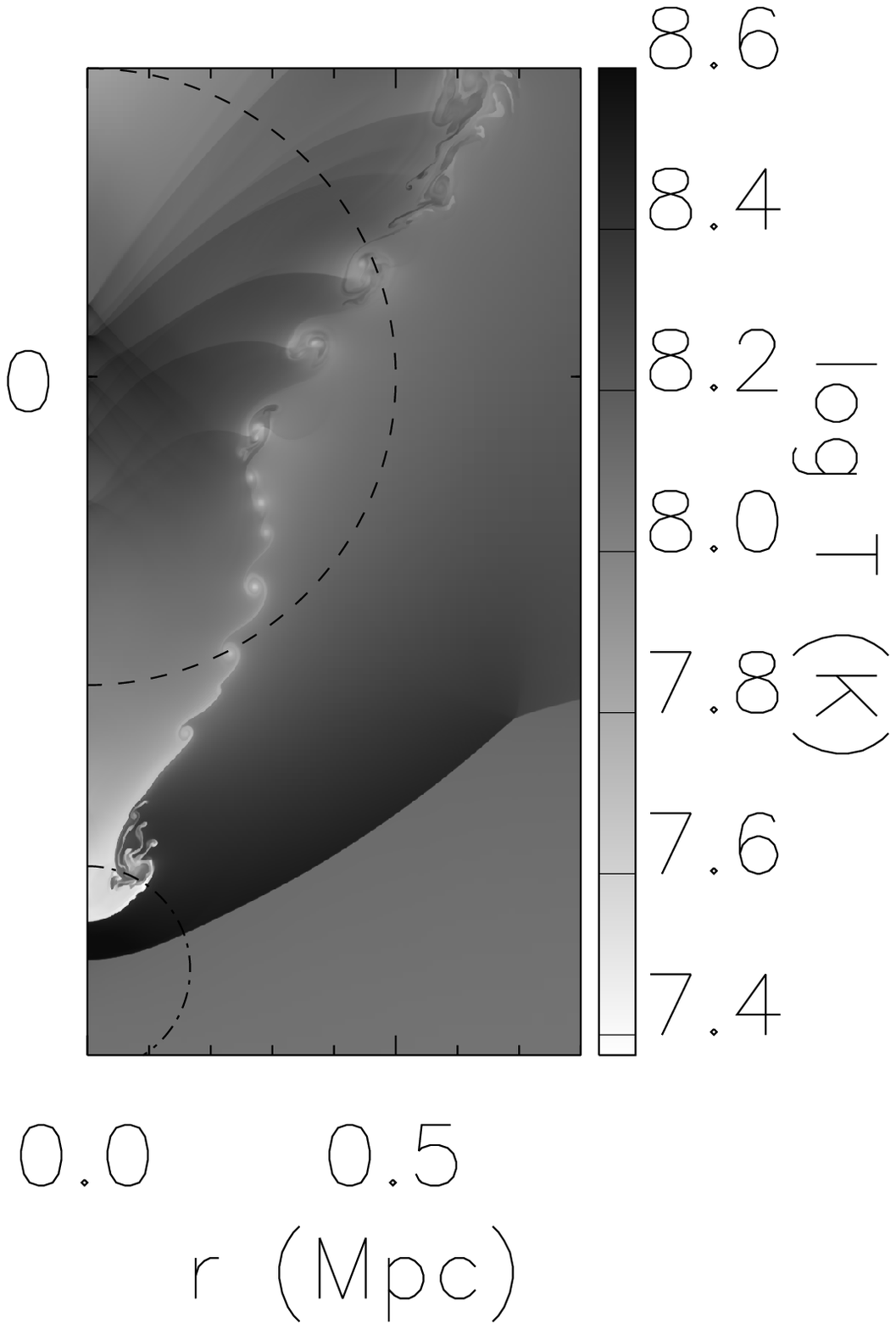}
      \end{center}
    \end{minipage}
  \end{tabular}
  \caption{Time evolution of temperature. Larger CDM halo is centered
    at cylindrical $(r,z)= (0, 0)$; smaller halo is at $z= 0.29,
    -0.66$ and $-0.96 \textrm{ Mpc}$, respectively. Circles indicate
    the scale radii $r_s$ of the halos.}
  \label{fig:temperature}
\end{figure}

\section{Results}
After subhalo ``bullet'' passes through the center of primary halo,
the subhalo gas trails the subhalo DM, led by a bow shock and contact
discontinuity (cold front)., as seen in Figure \ref{fig:temperature}.
 The
opening angle and radius of curvature of the bow shock are sensitive
to simulation details, but both are larger than expected for
steady-state bow shocks of hard spheres with constant velocity in a
uniform medium.  The wings of the contact discontinuity are
Kelvin-Helmholtz unstable.

\begin{figure}[bh]
  \begin{tabular}{ll}
    \begin{minipage}{0.52\textwidth}
      \begin{center}
	\includegraphics[width=8cm,clip]{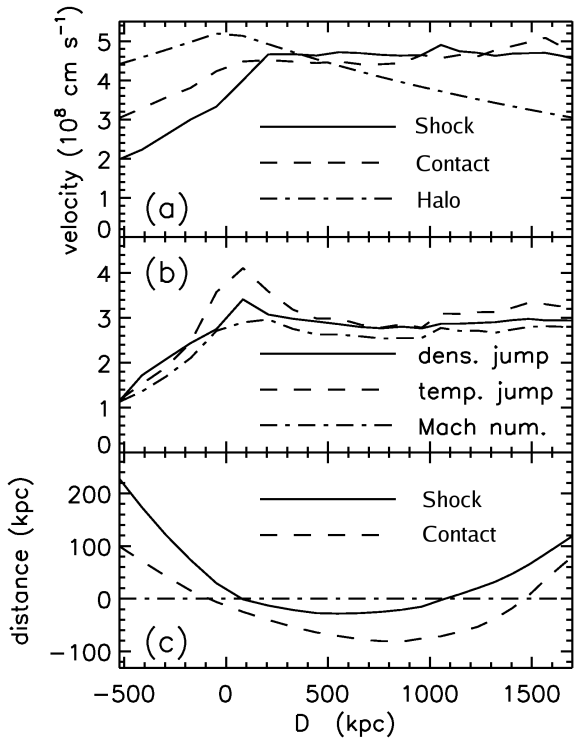}
      \end{center}
    \end{minipage}
    &
    \begin{minipage}{0.47\textwidth}
      \caption{(a) Velocity of the shock,
        $V_{\rm shock}$ (solid line), and of the contact discontinuity, $V_{\rm
          contact}$ (dashed line) relative to the preshock gas upstream, and the
        relative velocity of the CDM halos, $V_{12}$ (dash-dot line), all as functions
        of the time. The distance between two CDM halos, $D$, is
        used as the coordinate of time. (b) Density and temperature
        jump at the shock, and the shock Mach number.  (c) Position of
        the shock (solid line) and the contact discontinuity (dashed
        line) relative to subhalo center. Figures \ref{fig:shock} and 
        \ref{fig:temp-brightness} are taken from
        \citet{Milosavljevic:07}. 
        \vspace{25mm}
       }
      \label{fig:shock}
    \end{minipage}
  \end{tabular}
\end{figure}

\begin{figure}[bh]
  \begin{tabular}{lr}
    \begin{minipage}{0.31\textwidth}
      \begin{center}
	\includegraphics[height=41mm,clip]{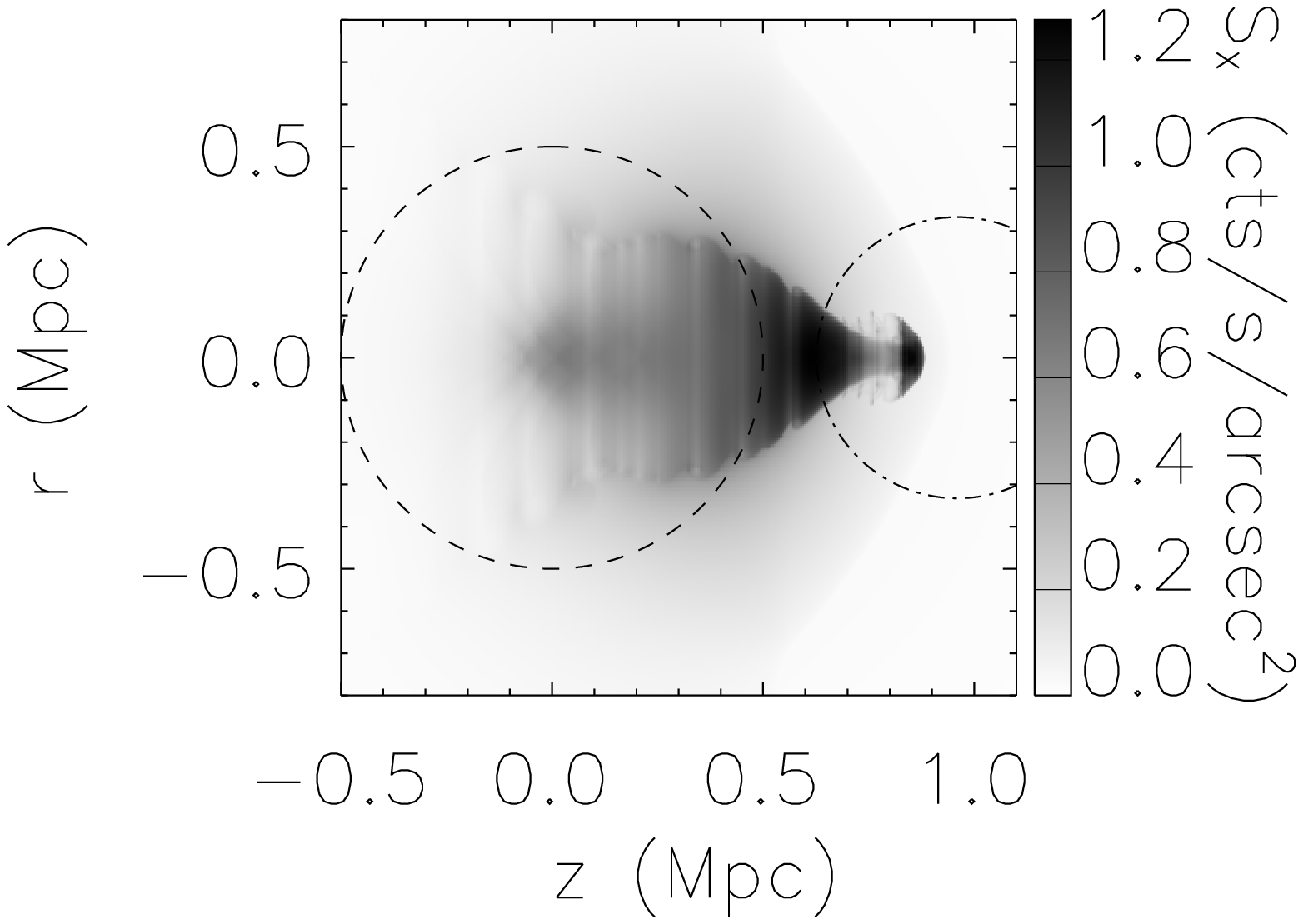}
      \end{center}
    \end{minipage}
    &
    \begin{minipage}{0.62\textwidth}
      \begin{center}
	\includegraphics[height=41mm,clip]{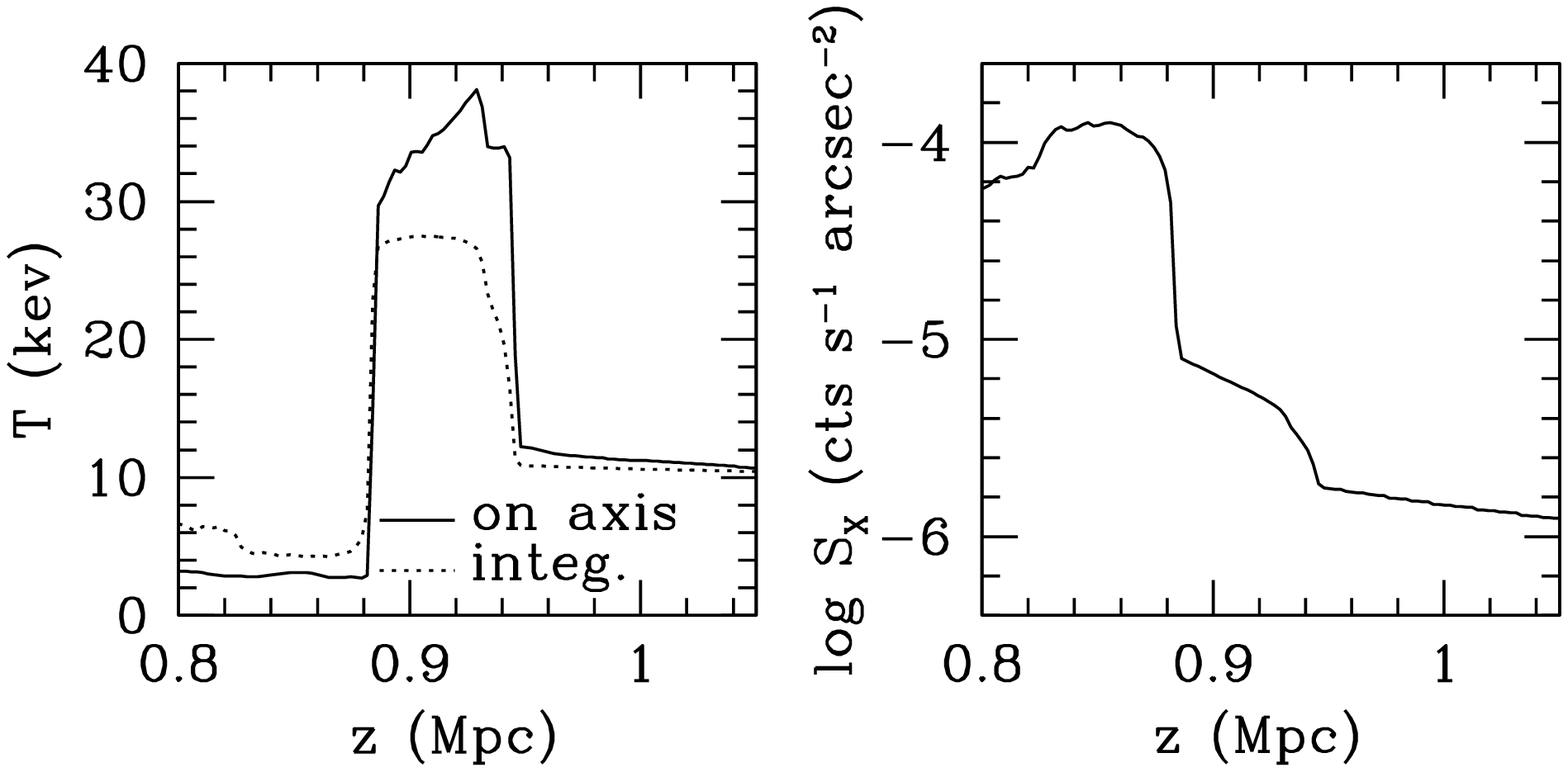}
      \end{center}
      
    \end{minipage}
  \end{tabular}

  \caption{ (Left:) Surface brightness map of the simulated cluster in
    the $0.8-4 \textrm{ keV}$ band. The circles indicate the scale
    radii $r_s$ of the CDM halos.  (Middle:) Temperature profile along
    collision axis (solid) and emissivity-weighted temperature,
    averaged over the (transverse) line of sight (dotted).  (Right:)
    Surface brightness profile along collision axis.
  }
  \label{fig:temp-brightness}
\end{figure}

During pericenter passage, the shock and contact
discontinuities are slower than the relative velocity of the two CDM
halos, due to ram pressure force acting on the gas. Later,
the halos climb out of the gravitational potential well and
decelerate, but the shock and contact discontinuity do not
decelerate (Fig. \ref{fig:shock}[a]).  At observed separation 
$D=720 \textrm{ kpc}$, the velocity of the shock is $4800 \textrm{ km s}^{-1}$,
consistent with the $4740 \textrm{ km s}^{-1}$ inferred from 1E
0657-56 data. The relative velocity of the simulated halos then is much
less, $4050 \textrm{ km s}^{-1}$ ($16\%$ smaller). 
The X-ray surface brightness and temperature profiles of the
simulation viewed transverse to collision axis
(Fig. \ref{fig:temp-brightness}) agree with those of 1E 0657-56.

\clearpage

\begin{figure}
  \begin{tabular}{cl}
    \begin{minipage}{0.45\textwidth}
      \begin{center}
	\includegraphics[height=6.0cm,clip]{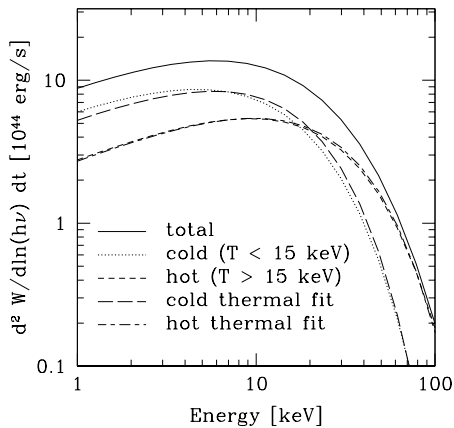}
      \end{center}
    \end{minipage}
    &
    \begin{minipage}{0.55\textwidth}
      \caption{Thermal bremsstrahlung emission spectrum from
        simulated merging cluster: the contribution from cold gas
        below $15 \textrm{ keV}$ (dotted) and hot gas above $15
        \textrm{ keV}$ (short-dash), total spectrum (solid), which is
        the sum of cold and hot components, and single-temperature fit
        to cold (long-dashed) and hot (mixed dash) components.  The
        emission from the hot shock-heated gas ($T>15 \textrm{ keV}$)
        dominates the spectrum above $20 \textrm{ keV}$ and is fitted
        well by a single temperature profile with $T = 20 \textrm{
          keV}$.  }
      \label{fig:spec}
    \end{minipage}
  \end{tabular}
  
\end{figure}

Figure \ref{fig:spec} shows the thermal X-ray spectrum of the
simulated merging cluster. The observed spectrum of 1E 0657-56 has
stronger emission above $30 \textrm{ keV}$, suggesting some additional
non-thermal emission \citep{Petrosian:06}.




\section{Discussion and Conclusions}
Our simulations of merging galaxy clusters show that the halo
collision velocity need not be the same as the intergalactic gas shock
velocity.  Applying our correction to the speed of the subhalo, the
probability that such a massive cluster in our ΛCDM universe has such
a high-speed subhalo increases by 3 orders of magnitude relative to
previous estimates, but is still small ($\sim 10^{-4}$).
\citet{Springel:07} independently reached a similar conclusion,
although their relative halo velocity is much smaller, $2600 \textrm{
  km s}^{-1}$. They assumed perfectly head-on collision between a
cuspy main halo and a more concentrated subhalo than ours, which would
make the subhalo decelerate more before reaching the observed
separation. More recent simulations by \citet{Mastropietro:07} show
that the difference between the shock velocity and the halo relative
velocity is smaller if the initial halo collision velocity is larger
and if the collision is not perfectly head-on. 




\begin{thebibliography}{}
\bibitem[Clowe et al.(2006)]{Clowe:06} Clowe, D., Brada{\v c}, 
M., Gonzalez, A.~H., Markevitch, M., Randall, S.~W., Jones, C., \& 
Zaritsky, D.\ 2006, \apjl, 648, L109 
\bibitem[Farrar \& Rosen(2007)]{Farrar:07} Farrar, G.~R., \& 
Rosen, R.~A.\ 2007, Physical Review Letters, 98, 171302 
\bibitem[Fryxell et al.(2000)]{Fryxell:00} Fryxell, B., et al.\ 
2000, \apjs, 131, 273 
\bibitem[Hayashi \& White(2006)]{Hayashi:06} Hayashi, E., \& 
White, S.~D.~M.\ 2006, \mnras, 370, L38 
\bibitem[Markevitch(2006)]{Markevitch:06} Markevitch, M.\ 2006, The 
X-ray Universe 2005, 604, 723 
\bibitem[Markevitch \& Vikhlinin(2007)]{Markevitch:07} Markevitch, 
M., \& Vikhlinin, A.\ 2007, Phys. Rep., 443, 1 
\bibitem[Mastropietro \& Burkert(2007)]{Mastropietro:07} Mastropietro, 
C., \& Burkert, A.\ 2007, preprint(arXiv:0711.0967)
\bibitem[Milosavljevi{\'c} et al.(2007)]{Milosavljevic:07} 
Milosavljevi{\'c}, M., Koda, J., Nagai, D., Nakar, E., \& Shapiro, P.~R.\ 
2007, \apjl, 661, L131 
\bibitem[Petrosian et al.(2006)]{Petrosian:06} Petrosian, V., 
Madejski, G., \& Luli, K.\ 2006, \apj, 652, 948 
\bibitem[Springel \& Farrar(2007)]{Springel:07} Springel, V., \& 
Farrar, G.~R.\ 2007, \mnras, 380, 911 
\end{thebibliography}
\end{document}